# Intrinsic dynamical fluctuations of PNS myelin

G. Campi[1], M. Di Gioacchino[1,2], N. Poccia[3], A. Ricci[2], M. Burghammer[4], A. Bianconi[1,2,5]

[1] Institute of Crystallography, CNR, via Salaria, Km 29.300, 00015 Monterotondo Roma, Italy
[2] Rome International Center for Materials Science Superstripes (RICMaSS), Via dei Sabelli 119A, 00185 Roma, Italy
[3] Department of Physics, Harvard University, Cambridge, Massachusetts 02138, USA∗
[4] European Synchrotron Radiation Facility, 6 Rue Jules Horowitz, BP220, 38043 Grenoble Cedex, France
[5] National Research Nuclear University MEPhI (Moscow Engineering Physics Institute) 115409 Moscow Kashirskoe shosse 31 Russia

**Abstract**

The ultrastructure fluctuations and complex dynamics of the multi-layered membrane structure of myelin are fundamental for understanding and control its formation process and its degeneration and repair in neurological diseases such as multiple sclerosis (MS). Myelin is considered a liquid-crystal but information are confined to its average structure due to limitations of the available standard techniques. To overcome this limitation in this work we have used Scanning micro X-ray Diffraction (SµXRD), which is a unique non-invasive probe of both k-space and real space allowing to visualize disorder in myelin with high spatial resolution in real space. We have used this method to examine the myelin sheath in the sciatic nerve of Xenopus laevis. Our results open could open new venues for understanding formation and degradation of myelin.

The compact myelin sheath is an elaborated multi-layered membrane that surrounds selected axons in the central and peripheral nervous systems (CNS, PNS) of vertebrates. It is fundamental for normal nervous system function[1]. Its main role is to decrease the capacitance of the membrane; it covers the axons, so that the speed of propagation of action potentials is increased via saltatory conduction, facilitating nerve signal transmission[2]. Myelin, in comparison with other biological membranes, have a very lipid-rich, dry mass, containing 75-80% of lipid and only 25-20% of proteins[3,4]. The major lipids in PNS myelin are cholesterol, phospholipids and galactolipids, which are important for membrane structure and assembly[1]. The major PNS compact myelin proteins are myelin basic protein, MBP, an unstructured protein, (present at only 5-18%[5]), the major structural protein of PNS myelin P0 glycoprotein[6], the peripheral myelin P2 and the peripheral myelin protein-22 (PMP-22). These proteins interact closely with the membrane[3].

Recently, there is increasing interest on the dynamics arising from these interactions, for a better understanding neurodegenerative diseases, such as multiple sclerosis[7,8], and nerve injuries[9]. Most of the studies have been focused on the dynamics of the atomic structure of individual proteins[3,10,11,12,13,14] and on understanding their structure and function[14,15,16,17,18,19,20]. Other studies have been focused on the influence of the proteins on the membrane dynamics, identifying the protein that control the fluctuations of the membrane layers thickness[3,11,15]. Also the myelin biogenesis has been studied, showing the dynamic development of the sheat[21,22,23] and of his chemical components and his metabolities[24]. There have been other studies on the dynamics of the diseases that incur on myelin[25,26] and on the dynamics of re-myelination and repair of the myelin sheath after a trauma[27,28]. In addition, we mention the change of the internal dynamics[29] as response to the variation of the myelin humidity. In this framework, there is interest and need to investigate the functional heterogeneity and the intrinsic structural fluctuations of myelin for a better understanding of myelination and demyelination processes and to discover the dynamic of supramolecular structure at nanoscale.

Myelin ultrastructure has been studied using a wide variety of approaches[10,24,30,31,32,33,34,35,36,37]; the most used are X-ray diffraction (XRD)[30-33], electron microscopy (EM)[34,35] and neutron diffraction[3,10,37]. The myelin ultrastructure can be seen as a multilamellar lattice with the repeating of a structural unity constituted by the stacking of the four following membranes: *i*) cytoplasmatic (*cyt*), *ii*) lipidic (lipid polar group, *lpg*), *iii*) extracellular (*ext*) and *iv*) another lipidic (*lpg*)[31,32,33]. This structural unit appears to be very stable because of intrinsic limitations of used experimental methods. Indeed, conventional XRD is limited since it provides insight only into the periodic structure of myelin probing the k-space (or reciprocal space) with no spatial resolution. On the other hand, electron microscopy approach (e.g. TEM) is a local, highly spatially resolved probe, but it is an invasive probe, which suffers from sample fixation and dehydration artefacts.

Therefore, we need a non-invasive probe with high spatial resolution to map functional spatial structure fluctuations. At this aim, we used Scanning micro X-ray Diffraction (SµXRD), which probes with high resolution both the k-space (reciprocal space) and real space inhomogeneity of complex materials; in a second step we applied tools of statistical physics to the diffraction collected data to unveil the *intrinsic structural disorder.*

This approach has been recently used to carry out a mapping of biological tissues with high spatial resolution, such as bone[38], cells[39] and myelin[18, 40].

We have measured the spatial disorder of the membranes thickness at thousands of discrete locations in freshly extracted and aged Xenopus sciatic nerves using a short time scale for data collection, obtained by using bright synchrotron radiation focused on 1µm$^2$ spot and the use of precise micro translation stages. Our results indicate while the periodicity of the myelin lattice show a *quasi-crystalline* state the lipidic hydrophobic layers and the cytoplasmatic and external hydrophilic layers show a large disorder.

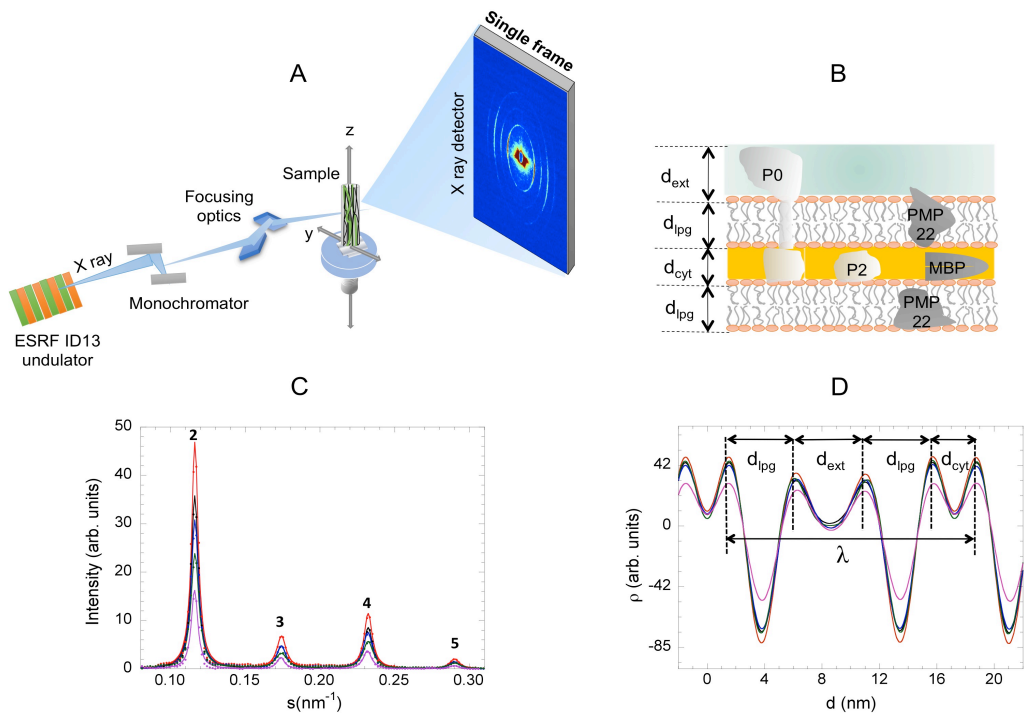

**Figure 1: A**) the SµXRD apparatus. **B**) Pictorial view of myelin periodic structure made of polar lipid groups, *lpg*, intercalated by two hydrophilic layers: the cytoplasm, *cyt*, and the extracellular apposition, *ext*. The specific myelin sheath protein PMP22, P0, P2 and MBP are indicated. **C**) Typical X-ray diffraction patterns measured at different sample spots. **D**) Electron density distribution computed presenting troughs and maxima, with a period λ=2$d_{lpg}$+$d_{ext}$+$d_{cyt}$, where $d_{lpg}$, $d_{ext}$ e $d_{cyt}$ refer to the thickness of lipid *lpg*, extracellular *ext*, cytoplasmatic *cyt* zones respectively.

The Scanning micro X ray Diffraction measurements of myelin of frog's sciatic nerve were performed on the ID13 beamline of the European Synchrotron Radiation Facility, ESRF, France. A scheme of the experimental setup is shown in Figure 1A. The source of the synchrotron radiation beam is a 18 *mm* period in vacuum undulator. The beam is first monochromatized by a liquid nitrogen cooled Si-111 double monochromator (DMC) and then is focused by a Kirkpatrick-Baez (KB) mirror system. This optics produces an energy X-ray beam of $\lambda$=12.6 *KeV* on a 1x1 $\mu m^2$ spot. The sample holder hosts the capillary-mounted nerve with the horizontal (y) and vertical (z) translation stages with 0.1 $\mu m$ repeatability.

The scheme of the multilamellar ultrastructure of myelin is shown in Figure 1B. It is made of the stacking of *i)* cytoplasmatic (*cyt*), *ii)* lipidic (*lpg*), *iii)* extracellular (*ext*) and *iv)* another lipidic (*lpg*)[31,32,33] layers. The individual thickness of each layer in 1 $\mu m^2$ area, named $\lambda$, $d_{cyt}$, $d_{lpg}$, $d_{ext}$, have been extracted from electron density profiles computed by Fourier analysis of the diffraction patterns.

The collected x-ray diffraction patterns measured at different sample spots are shown in Figure 1 **C**. Electron density distribution computed presenting troughs and maxima, with a period $\lambda=2d_{lpg}+d_{ext}+d_{cyt}$, where $d_{lpg}$, $d_{ext}$ e $d_{cyt}$ refer to the thickness of lipid *lpg*, extracellular *ext*, cytoplasmatic *cyt* zones respectively are shown in Figure 1D.

Our results demonstrate the feasibility of applying SµXRD for non-invasive imaging, to provide unique information on the disorder in biological systems and the data can be analysed using statistical physics approaches.[41-47] This technique allowed us to map the disorder in nanometric distribution of the myelin structural nanometer modules i.e., the cytoplasmatic, lipidic and extracellular subcomponents in the sciatic nerve of frog Xenopus leavis. From our measurements and from statistical analysis of distributions of these quantities we got the evidence that the *quasi-crystalline* periodicity of the myelin lattice is due to intrinsic disorder of hydrophobic and hydrophilic layers. Our results show that myelin structure, kept away from equilibrium in the biological system[48,49] shows typical disorder of the ultrastructure as in complex materials made of an assembly of nanoscale units[50-52]. Disorder in supramolecular assembly although it occurs widely in nature it is poorly understood since it takes place in multiple scales. Today it is of high relevance in health care, engineering[50] and photosynthetic processes[51]. The interests in modelling autonomous supramolecular assembly of materials is a vast challenging area in nanotechnology[52]. Therefore, wide varieties of supramolecular structures are possible depending on the nature of weak forces in biological liquid crystals like cholesterol and *myelin* ranging from *micro-* to *nano-* structures. For example, layered high temperature superconductors belong to this class of materials since they are composite systems resulting from the assembly of quantum wires or quantum wells[53,54] which have attracted large interest since they show the emergence of complex non Euclidean geometry that has been proposed to be a key factor promoting quantum coherence in living matter expected[55-57]. Finally, this work offers new perspectives for the study of emergence of neuro degenerative process and opens the way to further investigations into the ultrastructure dynamics, by studying time evolution of spatial fluctuations under external factors such as diseases and drug response.